# Evidence for superferrimagnetic clusters and spin-glass transition involving 4f Dy$^{3+}$ spins in h-DyMnO$_3$: A new twist to 4f Re$^{3+}$ spin ordering in hexagonal manganites

P. Aravinth Kumar,[1] Arun Kumar,[2] Keshav Kumar,[2] G. Anandha Babu,[1] P. Vijayakumar,[3] S. Ganesamoorthy,[3] P. Ramasamy[1] and Dhananjai Pandey[2]

[1] Department of Physics, SSN College of Engineering, Kalavakkam 603110, India

[2] School of Materials Science and Technology, Indian Institute of Technology (Banaras Hindu University), Varanasi 221005, India

[3] Materials Science Group, Indira Gandhi Centre for Atomic Research, Kalpakkam 603102, India

**Abstract:**

The ferroelectric phase of the multiferroic hexagonal manganites (h-ReMnO$_3$) has been reported to undergo a series of magnetic transitions involving long-range ordering/reorientation of 4fRe$^{3+}$ and/or 3dMn$^{3+}$ spins below room temperature. These transitions have attracted a lot of attention in recent years due to the geometrically frustrated nature of magnetic interactions. We have revisited these transitions in high quality single crystals of h-DyMnO$_3$ using dc and ac susceptibility measurements as a function of temperature (T), magnetic field (H) and frequency (ω) supplemented by specific heat measurements. Taking h-DyMnO$_3$ as an example, we show that the Dy$^{3+}$ spins below $T_N$~68K are in a superferrimagnetic (SFIM) state whereas they undergo spin-glass (SG) transition below $T_{Dy^{3+}}$~7K. Our observations demonstrate that neither the Néel transition at $T_N$~68K nor the transition at $T_{Dy^{3+}}$~7K is associated with long-range ordered states of Dy$^{3+}$ spins as believed so far in the literature. The SG state of h-DyMnO$_3$ is quite exotic as it occurs in an ordered compound purely due to geometrical frustration without any random disorder. Further, it shows an interesting crossover from de Almeida-Thouless type exponent (m=2/3) to Gabay-Toulouse type (m=2) with increasing field which cannot be explained in terms of the existing mean field theories of SG transition in Ising or Heisenberg systems but is expected for a vector X-Y SG system. Our observations call for a systematic reinvestigation of the nature of magnetic transitions involving $Re^{3+}$ ions in other h-$ReMnO_3$ also.

Recent years have witnessed enormous interest in multiferroic properties of rare earth manganites $ReMnO_3$ ($Re^{3+}$= $La^{3+}$-$Yb^{3+}$, $Y^{3+}$) which crystallise in the orthorhombic (o-) *Pnma* and hexagonal (h-) *P6$_3$cm* space group symmetries (SGS) for the larger and smaller $Re^{3+}$ ions, respectively [1-4]. For intermediate size $Re^{3+}$ ions like $Tb^{3+}$-$Lu^{3+}$, both the phases can be stabilized depending on the growth conditions (e.g., high pressure, inert or oxygen rich atmospheres etc…) [5-6]. Unlike the o-manganites which are type-II multiferroics, the h-manganites are type-I multiferroics with ferroelectric (FE) and antiferromagnetic (AFM) transition temperatures ($T_c$ and $T_N$) lying in the range 550-990K and 60-120K, respectively, depending on the $Re^{3+}$ ions [1,7-9]. The h-manganites, especially those with magnetic $Re^{3+}$ ions, have attracted considerable attention not only because of their strong magnetoelectric coupling but also because of the geometrically frustrated nature of the magnetic interactions that lead to highly degenerate states [10]. While both the $Mn^{3+}$ and $Re^{3+}$ spins in h-manganites are arranged on a triangular lattice, the former are always Heisenberg (non-collinear) type confined to the "ab" plane with in-plane frustrated magnetic interactions while the latter are believed to be Ising type aligned along the "c" axis [11,12]. The out-of-plane magnetic interactions between $Mn^{3+}$ spins as well as between $Mn^{3+}$ and $Re^{3+}$ spins are also frustrated in some of the h-manganites [13]. The AFM phase of most of the h-manganites below $T_N$ is reported to undergo a spin reorientation (SR) transition of the $Mn^{3+}$ spins at $T_{SR1}$~30 to 50K depending on the magnetic $Re^{3+}$ ion followed by two other transitions at still lower temperatures involving ordering/reordering of the $Re^{3+}$ spins at $T_{Re^{3+}}$~5 to 10K and a second $Mn^{3+}$ SR transition at $T_{SR2}$~2 to 3.5K [10-18]. All these magnetic transitions, including the Néel transition, are believed to lead to long-range ordered (LRO) phases.

Taking h-DyMnO$_3$ as example, we show here for the first time that the 4fDy$^{3+}$ spins are not long range ordered either below T$_N$~68K or below T$_{Dy}$$^{3+}$~7K. Among the h-manganites, the compound h-DyMnO$_3$ is rather special as Dy$^{3+}$ is the largest Re$^{3+}$ ion for which the h-phase has been stabilized at ambient pressure under inert gas atmosphere [6] without the application of high pressure conditions used for the synthesis of h-phase of Tb$^{3+}$-Lu$^{3+}$, Y$^{3+}$ manganites [5]. Further, recent x-ray resonant magnetic scattering (XRMS), second harmonic generation (SHG) and neutron diffraction (ND) studies suggest that the 3dMn$^{3+}$-4fRe$^{3+}$ exchange interactions in h-DyMnO$_3$ are less rigid as compared to other h-ReMnO$_3$ compounds [12,14] since the Dy$^{3+}$ and Mn$^{3+}$ spins in DyMnO$_3$ order in different magnetic SGS *P6$_3$cm* and *P6$_3$cm*, respectively, below T$_N$ giving an overall SGS as *P6$_3$* [19,20], in contrast to the other h-manganites (Re= Er$^{3+}$, Yb$^{3+}$, Tm$^{3+}$, Ho$^{3+}$ etc.) where both Re$^{3+}$ spins and Mn$^{3+}$ spins are found to order as per one irreducible representation of the *P6$_3$cm* SGS for the both the magnetic sublattices [12,14]. Further, these studies suggest that the AFM alignment of the Dy$^{3+}$ spins within the 2(a) and 4(b) sites below T$_N$~68K changes to FM type below T$_{Dy}$$^{3+}$~7K as a result of a spin-flop transition in the (001) plane but the spins at the two sites continue to maintain antiparallel alignment leading to an overall ferrimagnetic (FIM) state below T$_{Dy}$$^{3+}$~7K [19,20].

We have revisited the magnetic transitions in high quality single crystals of h-DyMnO$_3$ using dc and ac susceptibility measurements as a function of temperature (T), magnetic field (H) and frequency (ω) supplemented by specific heat measurements. We show here that the M-H plot becomes non-linear without any remanence (M$_r$) below T$_N$~68K and the non-linearity grows with decreasing temperature upto T$_{Dy}$$^{3+}$~7K suggesting that the spin-flop transition from the AFM state to the FIM state occurs only over short length scales leading to the formation of superferrimagnetic (SFIM) clusters of spins [21] and takes place gradually over a range of temperatures T$_{Dy}$$^{3+}$~7K<T<T$_N$~68K. More interestingly, we show that the $Dy^{3+}$ spins are not in the long-range ordered (LRO) FIM state even below $T_{Dy^{3+}} \sim 7K$ but enter into a spin-glass (SG) state as confirmed by bifurcation ($T_{irr}$) of the zero field cool (ZFC) and field cool (FC) susceptibility ($\chi_{dc}(T)$) curves at $T_{irr}>T_{Dy^{3+}}$, slow non-exponential relaxation of thermoremanent magnetization (TRM) M(t), memory and rejuvenation effect revealed through a "hole" burnt at the waiting temperature ($T_w$), critical slowing down of the spin dynamics confirming the non-ergodic nature of the SG phase below the SG transition temperature $T_{SG}$ =6.6K, $M_r$ below $T_{SG}$. The SG state of h-DyMnO$_3$ is quite exotic as it occurs in an ordered compound purely due to geometrical frustration without any random disorder [22,23]. Further, it shows an interesting crossover from de Almeida-Thouless (A-T) type exponent (m=2/3) to Gabay-Toulouse (G-T) type (m=2) with increasing field which cannot be explained in terms of the existing mean field theories of SG transition in Ising [24] or Heisenberg systems [25] but is expected for a vector X-Y SG system [26]. Our findings give a new twist to current understanding of the magnetic phase transitions in the family of h-manganites with magnetic $Re^{3+}$ ions where it has all along been assumed that $Re^{3+}$ spins are in the LRO state below $T_N$ and below $T_{Re^{3+}}$ and that these orderings induce SR transitions of the $Mn^{3+}$ spins below $T_{SR1}$ and $T_{SR2}$ through the rigid 3d$Mn^{3+}$-4f$Re^{3+}$ interactions [12-18]. We believe that our work will stimulate similar studies on other h-manganites also, as a proper understanding of the ordering scheme of the $Re^{3+}$ spins is vital to the understanding of the true ground state of h-ReMnO$_3$ compounds.

In the present study, we have used 4.5mm dia and 22mm length crystal of h-DyMnO$_3$ grown by optical floating zone technique. The details of sample preparation, crystal growth and characterizations are given in section S1 and S2 of the supplementary information (SI). It is intriguing to note that while upto four different magnetic transitions have been reported in various h-manganites with magnetic $Re^{3+}$ ion, no single experimental technique (e.g $\chi_{dc}(T)$ or specific heat ($C_p(T)$) or ND or SHG or XRMS) has revealed all the four transitions in most of these compounds including h-DyMnO$_3$ [6,19,20,27-29]. Our h-DyMnO$_3$ crystals show clear signatures of four different magnetic transitions in $\chi_{dc}(T)$ as well as C$_p$(T) measurements which demonstrates their excellent quality as compared to the crystals used by the previous workers [6,19,20,27-29]. Fig.1 of the main text and Fig.S3 of the SI depicts the $\chi_{dc}(T)$ measured under ZFC and FC conditions for a field H = 50 Oe applied approximately parallel (∥) and perpendicular (⊥) to the c axis, respectively. The anomaly around T$_{Dy^{3+}}$~7K in Fig.1 for H∥c is very prominent and much weaker for H⊥c. In addition, small anomalies corresponding to T$_N$~68K and the first SR transition at T$_{SR1}$~35K are also revealed as can be seen from the insets (a) and (b) of Fig.1. Further, a deviation from the trend of $\chi_{dc}(T)$ at low temperatures is also observed around T$_{SR2}$~2K (see inset (c) of Fig.1) for H∥c. The signature of the transitions at T$_{Dy^{3+}}$ and T$_{SR2}$ are also seen in $\chi_{dc}(T)$ for H⊥c (see insets (a) and (b) of Fig.S3 of SI). Our results are in agreement with the results of previous workers [6,20] for the anomalies at T$_N$ and T$_{Dy^{3+}}$ eventhough they did not observe any signature of T$_{SR1}$ and T$_{SR2}$. Mansouri et al. [27] have reported anomalies in $\chi_{dc}(T)$ at T$_{SR1}$ similar to that in inset (b) of Fig.1 but not at T$_{SR2}$. Thin film samples [29] also show evidence for transition at T$_{SR1}$ but they do not reveal the transition at T$_{Dy^{3+}}$. There is no previous report of the anomaly around T$_{SR2}$~2K in h-DyMnO$_3$ (shown in inset (c) of Fig.1 and inset (b) of Fig.S3) even though this transition has been reported in other h-manganites [12,13,15,17]. Our C$_p$(T) measurements shown in inset (e) of Fig.1 provide unambiguous confirmation of four transitions discussed above with transition temperatures matching within ±2K with those determined from $\chi_{dc}(T)$ plots. The anomaly at T$_{SR2}$~2K is depicted more clearly in the magnified view within the inset (e) of Fig.1. Previous workers reported evidence for only two transitions in C$_p$(T) corresponding to T$_N$ and T$_{Dy^{3+}}$ [6]. From the Curie-Weiss (C-W) plots (1/$\chi$(T) vs T), we obtain Curie-Weiss temperature $\theta_{CW}$= -39K and -4K for H∥ and ⊥ to c axis, respectively, confirming the AFM nature of the transition at T$_N$~68K.

The observation of all the four magnetic transitions in magnetic and specific heat studies confirms the excellent quality of the crystals used by us and we now focus on the transitions at $T_N$~68K and $T_{Dy^{3+}}$~7K which involve ordering/reordering of the $Dy^{3+}$ spins [6,19,20,27]. The magnetization measurements reveal linear M-H curve up to 9T for $T>T_N$ as expected for the paramagnetic phase. However, the M-H plots show non-linearity for both H∥c (see Fig.2 (a)) and H⊥c (see Fig.S4 of SI) without any measurable coercivity ($H_c$) or $M_r$ below $T_N$ suggesting the presence of short-range ordered (SRO) SFIM clusters of spins. The extrapolated value of the highest field (9T) magnetization to H=0 (i.e $M_r$) shows that it is higher for H∥c than that for H⊥c. We believe that the M(H=0) for H∥c and H⊥c are essentially due to $Dy^{3+}$ and $Mn^{3+}$ spins, respectively, as noted by previous workers also [20]. The non-linearity of the M-H plot grows with decreasing temperature as shown in the Fig.2 (a) and Fig.S4 of SI. Such a non-linearity of the M-H plot with SFIM characteristics below $T_N$ has not been reported so far in any of the h-manganites with magnetic $Re^{3+}$ ion [16]. Our results suggest that the spin-flop transition involving $Dy^{3+}$ spins at the 2(a) and 4(b) sites reported in XRMS and SHG measurements [19,20] does not occur abruptly at $T_{Dy^{3+}}$~7K but rather takes place over a broad range of temperatures ($T_{Dy^{3+}}<T<T_N$) and at short length scales only as indicated by zero $M_r$.

The ZFC and FC $\chi_{dc}(T)$ plots for H∥c show bifurcation just above $T_{Dy^{3+}}$~7K (see inset (d) of Fig.1). Further, the M-H loop opens up below $T_{Dy^{3+}}$~7K and shows small $M_r$ as well as $H_c$ (see Fig.2 (b)). Both the irreversibility of the FC and ZFC $\chi_{dc}(T)$ and non-zero $M_r$ can arise either due to blocking of SFIM clusters or a SG transition [21,30]. For non-interacting SFIM systems, the $H_c$ is known to follow the following temperature dependence $H_c = H_{c0}[(1-T/T_B)^{1/2}]$ [31], where $H_{c0}$ is the value of $H_c$ in the limit of T→0, below the blocking temperature. In our case, the plot of the $H_c$ vs $T^{1/2}$ is non-linear (see Fig.2 (c)) which rules out the SFIM blocking process. In order to further distinguish between SFIM blocking and SG freezing, we analysed the real ($\chi'(\omega, T)$) and imaginary ($\chi''(\omega, T)$) parts of the ac susceptibility $\chi(\omega, T)$ in the frequency range 33.3 to 9333.3Hz shown in Fig.1(f). It is evident from this figure that the temperature corresponding to the peak in both $\chi'(\omega, T)$ and $\chi''(\omega, T)$ shifts to higher side with increasing frequency of measurement as expected for both the SFIM blocking and SG freezing [21,30]. The Mydosh parameter $(K) = \Delta T/T_f[\Delta(log(\omega))]$, where $T_f$ is the $\chi'(\omega, T)$ peak temperature at the lowest frequency while $\Delta T$ is the difference between $\chi'(\omega, T)$ peak positions at two different frequencies, is often used to distinguish between blocking and SG freezing [32-33]. For h-DyMnO$_3$, $K$ obtained for the frequency range 33.3 to 9333.3Hz is 0.07 which lies in the expected range for SG transition and not SFIM blocking [32-33]. The Arrhenius plot (ln$\tau$ versus 1/T) for the relaxation time ($\tau$), obtained from the $\chi'(\omega, T)$ peak positions at various frequencies, shows non-linear behaviour (see Fig.3(a)) which also rules out the possibility of SFIM blocking [21]. On the otherhand, we could successfully model the temperature dependence of $\tau$ using Vogel-Fulcher (V-F) law $\tau = \tau_o exp(E_a/k_B(T-T_{VF}))$ [34,35] as well as the power law $\tau = \tau_o exp((T_f/T_{SG}) - 1)^{-zv}$ type critical dynamics [35,36], where $\tau_o$ is the inverse of the attempt frequency, $k_B$ the Boltzmann constant, $E_a$ the activation energy, $T_{VF}$ the V-F freezing temperature, $T_f$ the temperature corresponding to peak in the $\chi'(\omega, T)$ at each frequency ($\omega = 2\pi f$), $T_{SG}$ the SG transition temperature, z the dynamical scaling exponent and ν the critical exponent. The solid line in Fig.3(a) depicts the V-F fit while the straight line in the inset of this figure gives the power law fit. Both the fits are excellent and they give nearly similar values for the SG transition temperature $T_{VF} \sim T_{SG} \sim (6.7 \pm 0.2)$K and $\tau_o = (3.97 \pm 0.03) \times 10^{-13}$s. Our analysis thus shows that there exists a SG transition temperature $T_{SG}$ at which the slowest spin dynamics diverges in the limit of $\omega \rightarrow 0$ as a result of ergodic symmetry breaking. The low value of $\tau_o$ points towards canonical nature of the SG phase which was subjected various other tests as described below.

The peak temperature ($T_{max}$) in ZFC M(T) plot is known to shift to lower temperatures along the A-T line on increasing the magnetic field (H) as a result of replica symmetry breaking following the relationship $H^2=A[\{T_G-T_{max}(H)\}/T_G]^3$ for low fields in SG systems [24]. In the present system also, the $T_{max}$(H) decreases with increasing H (see Fig.S5 of SI which depicts ZFC $\chi_{dc}$(T) at different fields for H∥c) and that the plot of $T_{max}$(H) against $H^{2/3}$ is linear for low fields (H<1000 Oe) as expected for the A-T line (see Fig.3(b)) for a model SG system [32-33]. The extrapolation of the A-T line to H=0 gives $T_{SG}$=6.6K which is in excellent agreement with $T_{SG}$ determined from $\chi(\omega, T)$ measurements. The $T_{max}$ vs $H^{2/3}$ plot deviates from linear behaviour for H>1000 Oe. Interestingly, $T_{max}$ shows $H^2$ dependence (see inset of Fig. 3(b)) above H>1000 Oe as expected for the G-T line [25]. This crossover from A-T type exponent (m=2/3) to G-T type (m=2) is quite interesting, as it is not consistent with Ising nature of $Dy^{3+}$ spins (or $Re^{3+}$ spins in general) as believed so far [24]. While a crossover between A-T and G-T lines has been known for disordered Heisenberg systems with increasing field or decreasing temperature, this would have required existence of two SG transitions as a result of freezing of longitudinal and transverse components of the SG order parameter in h-DyMnO$_3$ because of its low but positive value of D/J, where D is the single ion anisotropy and J is the exchange interaction parameter [25]. The existence of only single SG transition with such a crossover cannot therefore be rationalised in terms of the existing mean field theories of SG transition either in Ising or Heisenberg systems. In a very recent work [26], it has been shown numerically that A-T to G-T type crossover can occur in vector X-Y SG systems as a function of increasing magnetic field which modifies the local field experienced by the spins from being random to uniform. We believe that the crossover observed in h-DyMnO$_3$ could be due to freezing of the transverse component of the $Dy^{3+}$ spins leading to vector X-Y glass [26].

Additional confirmation for the SG state below $T_{Dy}^{3+}$ temperature was obtained by the observation of memory and rejuvenation effect below $T_{SG}$~6.6K reported in cluster glass and atomic SG systems [21,37] and explained in terms of the chaotic ground state [38]. To illustrate this, we depict in Fig.3(c) the ZFC M(T) data recorded at 100 Oe with and without intermediate stop at waiting temperature ($T_w$)=4K below $T_{SG}$ for a wait time $t_w$ of $10^4$s. The $\Delta M^{ZFC}(T) = (M_{wait}^{ZFC}(T)-M_{ref}^{ZFC}(T))$ vs T plot shown in Fig.3(c)) reveals a sharp "hole burning" dip exactly at $T_w$ which confirms the memory and rejuvenation effect in the SG phase of h-$DyMnO_3$. We also verified slow relaxation of the TRM below $T_{SG}$ observed in SG system [39-40]. For this, the sample was first cooled from 300 to 4K in a field of 100 Oe. Then after waiting for $t_w$=500s at 4K, the field was removed and the slow decay of the TRM was recorded as a function of time. The observed decay of TRM, shown in Fig.3(d), follows stretched exponential behaviour *M(t)=$M_r$+$M_G$ exp[-(t/τ)$^{(1-n)}$]*, where $M_r$ is the remanent ferromagnetic component, $M_G$ the glassy component, τ the characteristic relaxation time and n the stretched exponential exponent [39-40]. The best fitted curve shown in Fig.3d correspond to $M_r$=1.17 emu/g, $M_G$=0.014 emu/g, n=0.42 and τ=2252 s for $t_w$=500s. The value of exponent n is in agreement with the values reported in the literature for SG systems [39-40].

To summarise, our results demonstrate that the AFM to FIM spin-flop transition occurs over a broad range of temperatures $T_{Dy}^{3+}$~7K<T<$T_N$~68K and not abruptly at $T_{Dy}^{3+}$~7K as believed so far [6,19,20,27]. Further this transition occurs only at short length scales leading to the formation of SFIM clusters of spins not reported in any of the h-manganites. More significantly, we have discovered a SG transition around $T_{Dy}^{3+}$~7K not reported in any one of the h-$ReMnO_3$ compounds including h-$DyMnO_3$ [6,19,20,27] so far. This SG phase is unique as it occurs in an ordered compound due to geometrical frustration only and shows a crossover from A-T to G-T line at high magnetic fields. While recent years have witnessed some advancement in the theoretical understanding of SG transition in geometrically frustrated ordered systems [23,26], the SG phase of h-DyMnO$_3$ is more complex because it occurs in a ferroelectric state that not only makes the exchange interactions asymmetric but also leads to coupling of the various ferroic order parameters (ie, magnetoelastic and magnetoelectric couplings) as demonstrated recently in the SG phases of disordered multiferroic BiFeO$_3$ [41]. We believe that our observations will stimulate further experimental, especially using microscopic probes like neutron scattering and Mössbauer spectroscopy [37,41-42], and theoretical investigations on h-ReMnO$_3$ manganites, in general, and h-DyMnO$_3$ in particular.

**Acknowledgments:** This work is supported by UGC-DAE-CSR, Kalpakkam Node under project No. CSR-KN/CRS-46/2013-2014/646. Authors are thankful to Dr. A. Thamizhavel, TIFR for checking the orientation of the crystals.

**Figure Captions:**

**Fig. 1.** Temperature dependence of the dc susceptibility of h-DyMnO$_3$ for H∥c axis measured under ZFC (■) and FC (●) conditions at H=50 Oe. Insets **(a-d)** depict the magnified view near the three transition temperatures and irreversibility. **(e)** Shows C$_p$ vs T plot, while inset in this figure depicts the signature of 2K transition on a magnified scale. **(f)** depicts temperature dependence of the real and imaginary parts of the χ (ω,T) at various frequencies (33.3(■), 333.3(●), 1333.3(▲), 3333.3(♦), 5333.3(▼), 7333.3(∗) and 9333.3 Hz(◄)) for h-DyMnO$_3$ along c axis.

**Fig. 2(a)** Isothermal magnetisation curves of h-DyMnO$_3$ along the c axis measured at different temperatures, T=1.8K(■), 2K(●), 3K(▲), 5K(▼), 7K(◄), 10K(►), 15K(♦), 20K(□), 25K(○), 30K(△), 35K(▽), 40K(◁), 50K(▷), 60K(∗), 65K(☆), 70K(Θ), 80K(⬠) and 90(1)K, **(b)** shows the M-H hysteresis loop below $T_{SG}$~6.6K and **(c)** shows $H_c$ versus $T^{1/2}$ plot below 10K.

**Fig. 3(a)** Shows nonlinear nature of ln τ versus 1/T plot. The solid line shows the fit for Vogel-Fulcher law while the inset shows fit for power law dynamics, **(b)** depicts the field dependence of $T_{max}$. The solid line is the fit for de Almeida-Thouless criticality while the inset shows the fit for Gabay-Toulouse criticality, **(c)** shows a dip corresponding to ΔM (($M_{ref}^{ZFC}$)-($M_{wait}^{ZFC}$)) vs T plot obtained from H=100 Oe at waiting temperature 4K and **(d)** shows the variation of thermoremanent magnetization with time for h-DyMnO$_3$ and solid line is fit for stretched exponential function. All measurements are along the c-axis.

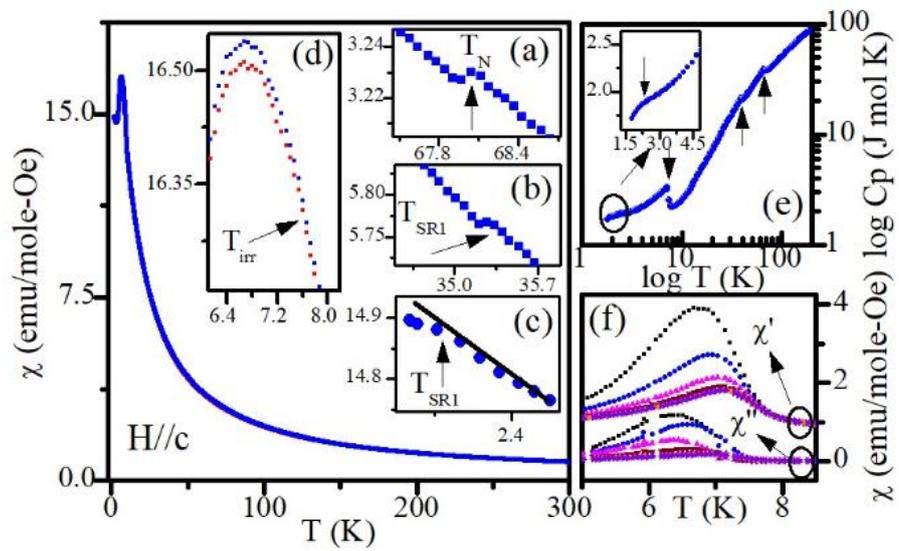

**Figure 1.**

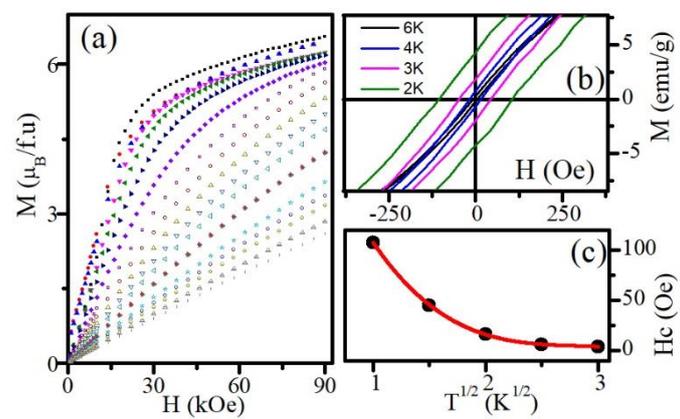

**Figure 2.**

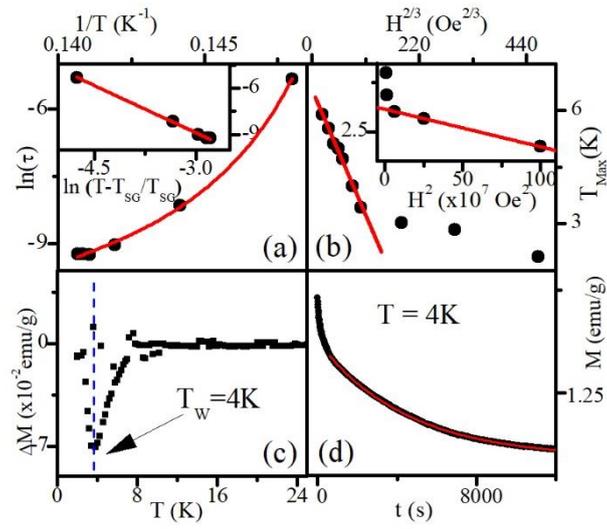

**Figure 3.**

# Supplementary Information

## S1. Experimental details:

DyMnO$_3$ was synthesized by solid state route using high purity oxides of Dy$_2$O$_3$ and Mn$_2$O$_3$ (Alfa Aeasr, 99.99%) as starting materials. Initially Dy$_2$O$_3$ is heated at 800 ºC in air for 12 h to remove any moisture. The precursors were taken in stoichiometric proportions and mixed for 2 h. The mixed powder was then heated at 850 ºC for 24 h in a closed platinum crucible. Further, the heated material was crushed and again fired at 1250 ºC for 24 h. The resulting material was again crushed into a fine powder to check the phase formation using X-ray powder diffractometer. X-ray powder diffraction pattern clearly confirm the single-phase formation of o-DyMnO$_3$.

The powder so obtained was used to stuff it in a rubber tube to make feed rod. The filled rubber tube was first evacuated, and then sealed and finally pressed in a cold isostatic at a pressure of about 70 MPa. The pressed compact rod was carefully separated from rubber tube and sintered at 1200 ºC for 24 h. The sintered rods were used for growing single crystals using optical floating zone furnace with four halogen lamps and ellipsoidal mirrors (Crystal system Corp. FZ-T-4000-H-HR-I-VPO-PC). Growth runs were performed under high purity argon (Ar) atmosphere at a pressure of 2 bar. A number of growth experiments were performed to optimise the parameters like sample rotation speed and growth rate.

It was verified that the structure of the DyMnO$_3$ crystals grown under oxygen ambience remained orthorhombic whereas it transformed to hexagonal structure when the crystals were grown in Ar atmosphere. This is in agreement with previous reports [1]. Fig. S1. shows the image of one such as grown crystal.

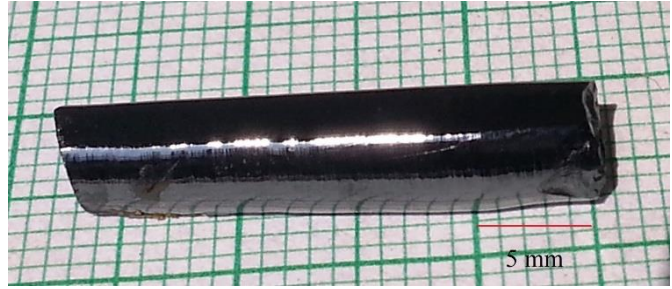

**Fig. S1** Photograph of as grown h-DyMnO$_3$ single crystal.

**S2. Characterizations**

X-ray powder diffraction (XRPD) pattern was recorded at room temperature using a STOE diffractometer operated in Bragg–Brentano geometry with fixed slits. For this, we crushed a small portion of the grown crystal into fine powder and XRPD pattern was recorded in the 2θ range 20-90º with a step size of 0.05º. LeBail refinement was performed by the FULLPROF package [2] for checking the phase purity. The dc magnetization [M (T, H)] measurements were carried out using a physical property measurement system (PPMS, Quantum Design) in the temperature range 2-300K. The ac magnetic susceptibility was carried out on the same system at various frequencies ranging from 33.3 Hz to 9333.3 Hz as function of temperature. The specific heat measurements were also performed on the same system in the temperature range 2-200K.

**S3. Room temperature crystal structure:**

Fig. S2. depicts the observed (filled circles), calculated (continuous line) and difference (bottom line) profiles obtained from LeBail refinement. The satisfactory fits between the observed and calculated profiles confirms that the grown crystal belongs to hexagonal crystal structure in the *P6$_3$cm* space group. The refined lattice parameters a = 6.174025 (1) Å and c = 11.452806 (3) Å are in perfect agreement with the previous workers [1]. No evidence for the orthorhombic phase was observed in the XRPD patterns.

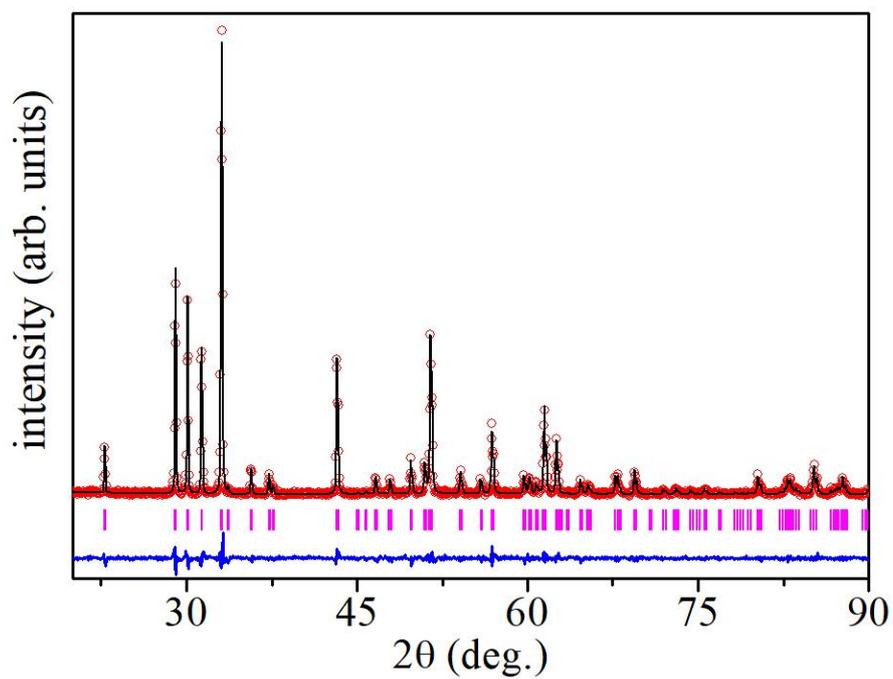

**Fig. S2.** The observed (red open circles), calculated (black continuous line) and difference (bottom blue line) profiles obtained from LeBail refinement using *P6₃cm* space group at room temperature. The vertical lines (pink) indicate the Bragg peak positions for h-DyMnO$_3$.

## S4. Figures referred in the main text

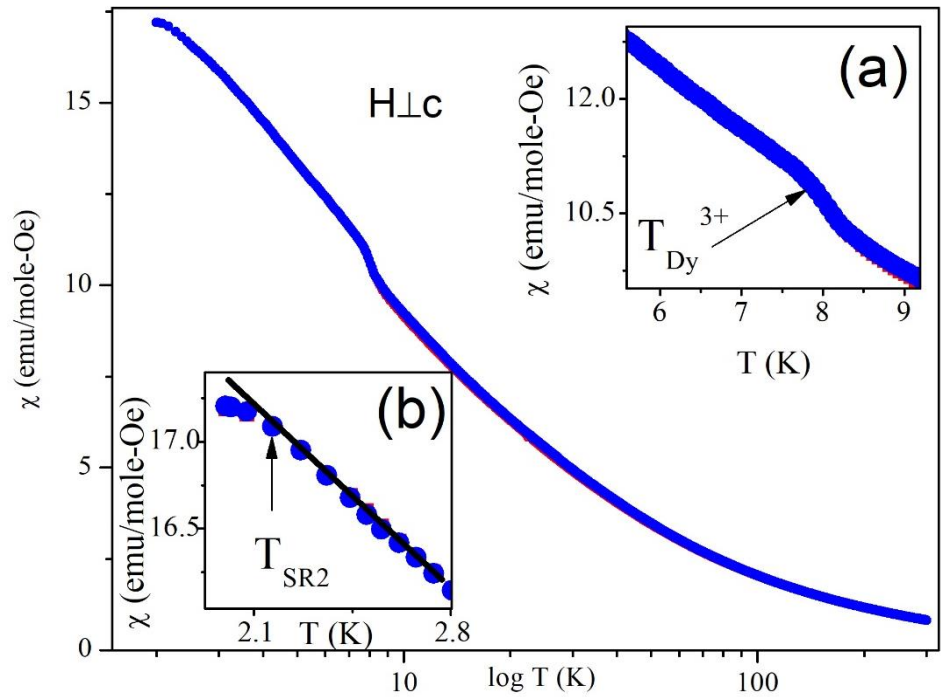

**Fig. S3** Temperature dependence of the dc magnetization measured under ZFC (■) and FC (●) conditions with a applied field of 50 Oe perpendicular to c axis for h-DyMnO$_3$. Insets depict the magnified view (a) around T$_{Dy^{3+}}$~7K and (b) near T$_{SR2}$ transitions.

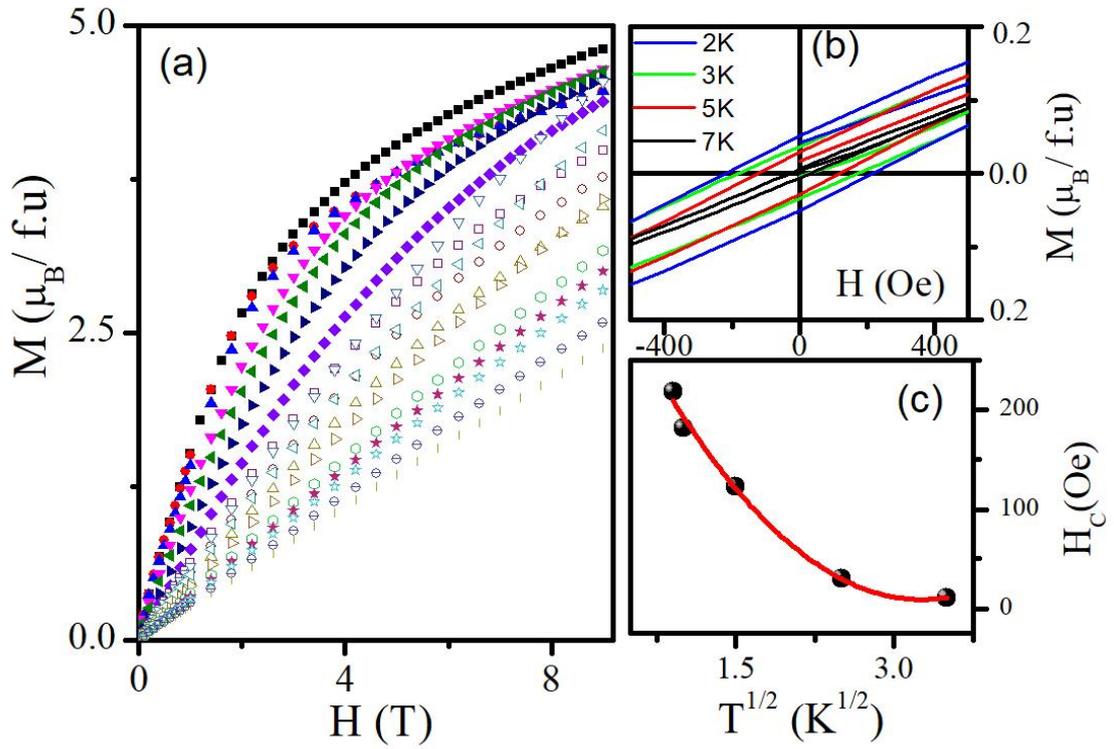

**Fig. S4(a)** Isothermal magnetization curves of h-DyMnO$_3$ perpendicular to the c axis measured at different temperatures, T =1.8K(■), 2K(●), 3K(▲), 5K(▼), 7K(◄), 10K(►), 15K(♦), 20K(□), 25K(○), 30K(∆), 35K(∇), 40K(◁), 50K(▷), 60K(∗), 65K(☆), 70K(Θ) 90K(|), Panel **(b)** depicts the M-H hysteresis loop below 10K while **(c)** shows the H$_c$ versus T$^{1/2}$ plot below 10K.

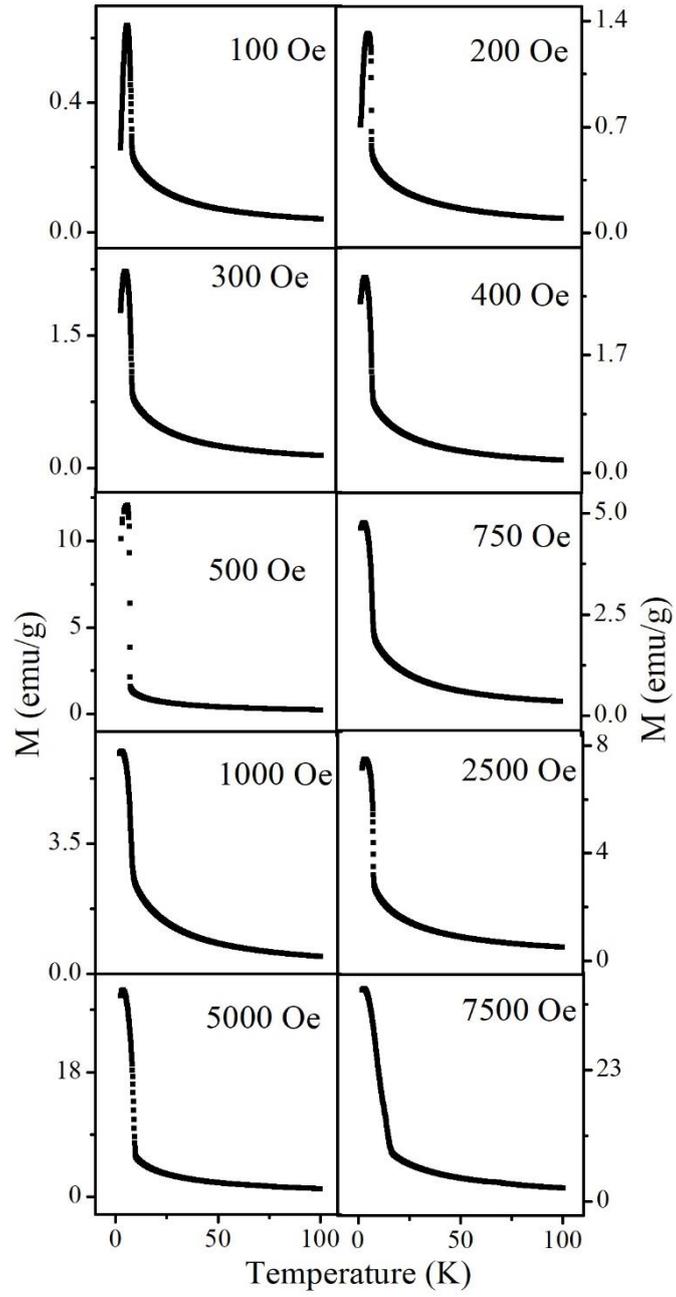

**Fig. S5** Temperature dependence of the dc susceptibility of h-DyMnO$_3$ for H∥c axis measured under ZFC condition with different magnetic fields.

**References for supplementary information:**